\documentclass[manuscript]{aastex631}

\usepackage{mathtools}
\usepackage{threeparttable}
\shorttitle{Observational Evidence of Plasmoid Instability}
\shortauthors{Lu et al.}
\graphicspath{{./}{figures/}}
\begin{document}

\title{Observational Signatures of  Tearing Instability in the Current Sheet of a Solar Flare}

\correspondingauthor{Li Feng}
\email{lfeng@pmo.ac.cn}

\author[0000-0002-3032-6066]{Lei Lu}
\affiliation{Key Laboratory of Dark Matter and Space Astronomy, Purple Mountain Observatory, Chinese Academy of Sciences, 210033 Nanjing, People's Republic of China}
\affiliation{CAS Key Laboratory of Solar Activity, National Astronomical Observatories, Beijing 100012, China}

\author{Li Feng}
\affiliation{Key Laboratory of Dark Matter and Space Astronomy, Purple Mountain Observatory, Chinese Academy of Sciences, 210033 Nanjing, People's Republic of China}
%
%\collaboration{6}{(AAS Journals Data Editors)}
%
\author{Alexander Warmuth}
\affiliation{Leibniz-Institut f\"{u}r Astrophysik Potsdam (AIP), An der Sternwarte 16, D-14482 Potsdam, Germany}

\author{Astrid M. Veronig}
\affiliation{Institute of Physics \& Kanzelh\"ohe Observatory for Solar and Environmental Research, University of Graz, A-8010 Graz, Austria}

\author{Jing Huang}
\affiliation{CAS Key Laboratory of Solar Activity, National Astronomical Observatories, Beijing 100012, China}

\author{Siming Liu}
\affiliation{School of Physical Science and Technology, Southwest Jiaotong University, Chengdu 610031, China}

\author{Weiqun Gan}
\affiliation{Key Laboratory of Dark Matter and Space Astronomy, Purple Mountain Observatory, Chinese Academy of Sciences, 210033 Nanjing, People's Republic of China}

\author{Zongjun Ning}
\affiliation{Key Laboratory of Dark Matter and Space Astronomy, Purple Mountain Observatory, Chinese Academy of Sciences, 210033 Nanjing, People's Republic of China}

\author{Beili Ying}
\affiliation{Key Laboratory of Dark Matter and Space Astronomy, Purple Mountain Observatory, Chinese Academy of Sciences, 210033 Nanjing, People's Republic of China}

\author{Guannan Gao}
\affiliation{Yunnan Observatories, Chinese Academy of Sciences,  Kunming 650011, China}

%% Note that the \and command from previous versions of AASTeX is now
%% depreciated in this version as it is no longer necessary. AASTeX 
%% automatically takes care of all commas and "and"s between authors names.

%% AASTeX 6.31 has the new \collaboration and \nocollaboration commands to
%% provide the collaboration status of a group of authors. These commands 
%% can be used either before or after the list of corresponding authors. The
%% argument for \collaboration is the collaboration identifier. Authors are
%% encouraged to surround collaboration identifiers with ()s. The 
%% \nocollaboration command takes no argument and exists to indicate that
%% the nearby authors are not part of surrounding collaborations.

%% Mark off the abstract in the ``abstract'' environment. 
\begin{abstract}

Magnetic reconnection is a fundamental physical process converting magnetic energy into not only plasma energy but also particle energy in various astrophysical phenomena.
In this letter, we show a unique dataset of a solar flare where various plasmoids were formed by a continually stretched current sheet. 
EUV images captured reconnection inflows, outflows, and particularly the recurring plasma blobs (plasmoids).
X-ray images reveal nonthermal emission sources at the lower end of the current sheet, presumably as large plasmoids with a sufficiently amount of energetic electrons trapped in.
In the radio domain, an upward slowly drifting pulsation structure, followed by a rare pair of oppositely drifting structures, was observed. These structures are supposed to map the evolution of the primary and the secondary plasmoids formed in the current sheet. Our results on plasmoids at different locations and scales shed important light on the dynamics, plasma heating, particle acceleration, and transport processes in the turbulent current sheet and provide observational evidence for the  cascading magnetic reconnection process.

\end{abstract}

%% Keywords should appear after the \end{abstract} command. 
%% The AAS Journals now uses Unified Astronomy Thesaurus concepts:
%% https://astrothesaurus.org
%% You will be asked to selected these concepts during the submission process
%% but this old "keyword" functionality is maintained in case authors want
%% to include these concepts in their preprints.

%% From the front matter, we move on to the body of the paper.
%% Sections are demarcated by \section and \subsection, respectively.
%% Observe the use of the LaTeX \label
%% command after the \subsection to give a symbolic KEY to the
%% subsection for cross-referencing in a \ref command.
%% You can use LaTeX's \ref and \label commands to keep track of
%% cross-references to sections, equations, tables, and figures.
%% That way, if you change the order of any elements, LaTeX will
%% automatically renumber them.
%%
%% We recommend that authors also use the natbib \citep
%% and \citet commands to identify citations.  The citations are
%% tied to the reference list via symbolic KEYs. The KEY corresponds
%% to the KEY in the \bibitem in the reference list below. 

\section{Introduction} \label{sec:intro}

Magnetic reconnection is a fundamental process in plasma physics that is relevant not only in the context of solar and stellar flares, but also in planetary magnetospheres, magnetars, accretion disks, and in laboratory plasmas \citep{Priest2000,Lin2008,Zweibel2009}. It was first proposed and has been widely used to explain the energy release in solar eruptions \citep{Su2013,Li2016,Cairns2018,Gou2019}. In the classical CSHKP model \citep{Carmichael1964, Sturrock1966,Hirayama1974,Kopp1976}, a closed magnetic structure is stretched by a rising flux rope (usually observed as a filament), forming a current sheet (CS)  where oppositely directed magnetic field lines flow in and reconnect. The energy pre-stored in the magnetic field is then converted into various energy forms such as heating of plasma, acceleration of particles, bulk mass motions, and emissions in almost all wavelengths. The newly formed field lines retract both downward and upward from the reconnection site, forming post-flare loops and coronal mass ejections (CMEs).
The CSHKP model agrees well with the large-scale dynamics of the observed eruptive events. However, 
the reconnection rate estimated from the model is often found to be too low to explain the rapid energy release in solar flares \citep{Shibata2011}.
Moreover,  with a single diffusion (reconnection) region, the CSHKP model also shows an apparent inability to account for the acceleration of the  observed large number of  energetic particles \citep{Fletcher2005,Krucker2008}.

Because of these difficulties, a scenario of cascading reconnection was suggested by \cite{Shibata2001}. 
In this scenario, the initial CS is continually stretched  by the rising flux rope (CME), at some point the tearing-mode instability sets in and  multiple magnetic islands (also called plasmoids) are formed, interleaved with thin CSs. 
The newly formed plasmoids are subjected to increasing separation,  leading to the secondary tearing instability that causes the  CSs to be further filamented. 
As the process continues, third and higher orders of tearing instabilities take place and plasmoids with smaller and smaller size are formed until the widths of the CSs reach the plasma-kinetic scales at which the magnetic energy is dissipated.
This scenario was later supported by the 
analytical theory of chain plasmoid instability \citep{Loureiro2007,Uzdensky2010} and confirmed by various numerical simulations \citep{Samtaney2009,Bhattacharjee2009, Huang2010,Shen2011,Barta2011,Mei2012,Ni2015, Mei2017,Lid2019,Zhao2021}.
It has been shown that the reconnection rate as well as the acceleration of solar particles can be significantly enhanced by the formation and ejection of the secondary and higher orders of plasmoids.

These theoretical predictions are consistent with  laboratory modelling of  secondary CSs  in laser-plasma interaction \citep{Dong2012} and in situ observations of a secondary magnetic island in the Earth's magnetotail \citep{Wang2010}. 
In solar eruptive events, the plasmoids were first recognized from soft X-ray images  during the impulsive phase of a solar flare \citep{Shibata1995}. Later, they were also identified in hard X-ray, extreme ultraviolet (EUV), white light, and radio images \citep{Hudson2001,Ko2003,Shimojo2017}.  Additionally, drifting pulsation structures (DPSs) which are sometimes observed in the solar radio dynamic spectra are also interpreted as signatures of  plasmoids \citep{Kliem2000,Karlicky2004,Liu2010}. 
However, the formation of DPSs is a very complex process that requires many specific conditions. They can only be observed in some special cases. 
In this Letter, we present a unique dataset where the plasmoid formation and evolution in a turbulent current sheet during magnetic reconnection are directly observed in unprecedented detail.

\section{Observations and analysis} \label{sec:obs}

\begin{figure}[ht!]
\plotone{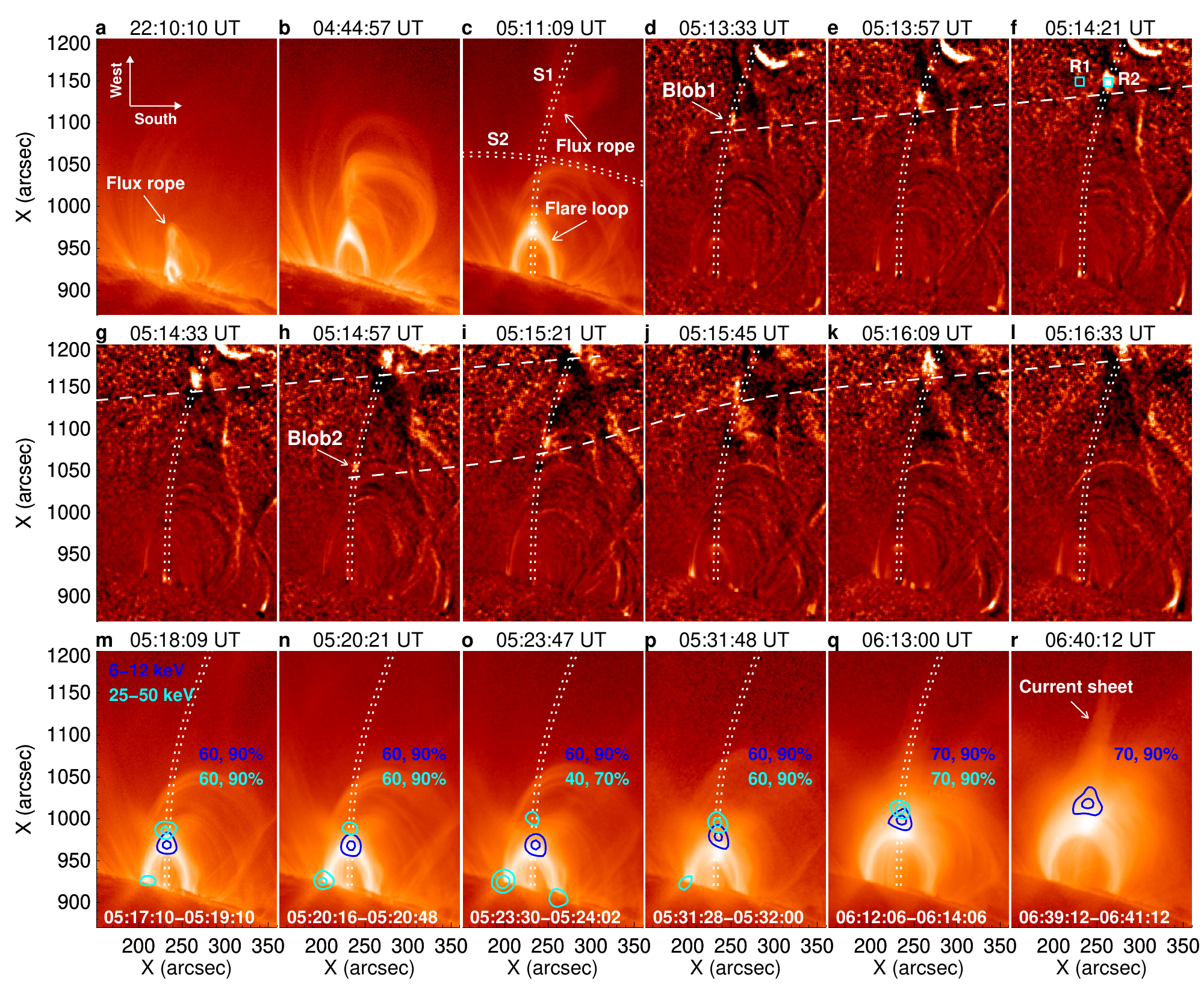}
\caption{Snapshots of the eruption recorded by the 131 \AA ~ filter of SDO/AIA. West is rotated to the top and south to the right.
Panel(a) shows the flux rope formed in an earlier confined C4.5 flare. Panels (b-r) show the eruption and evolution of the flux rope and the associated M7.7 flare. ``S1" and ``S2" in panel (c) indicate two slits placed along and perpendicular to the current sheet.
Panels (d-l) are running ratio images, showing motions of two blob-like plasmoids.
``R1''and ``R2'' in panel (f) indicate two subregions used for the  DEM analysis.
The arrow in panel (r) refers to the long current sheet. 
The contours overplotted in panels (m)-(r) show the RHESSI X-ray sources at 6-12 keV (blue) and 25-50 keV (cyan) reconstructed with the ``clean'' algorithm.
The percentages define the contour levels.
\label{fig:blob_img_evo}}
\end{figure}

\begin{figure}[ht!]
\plotone{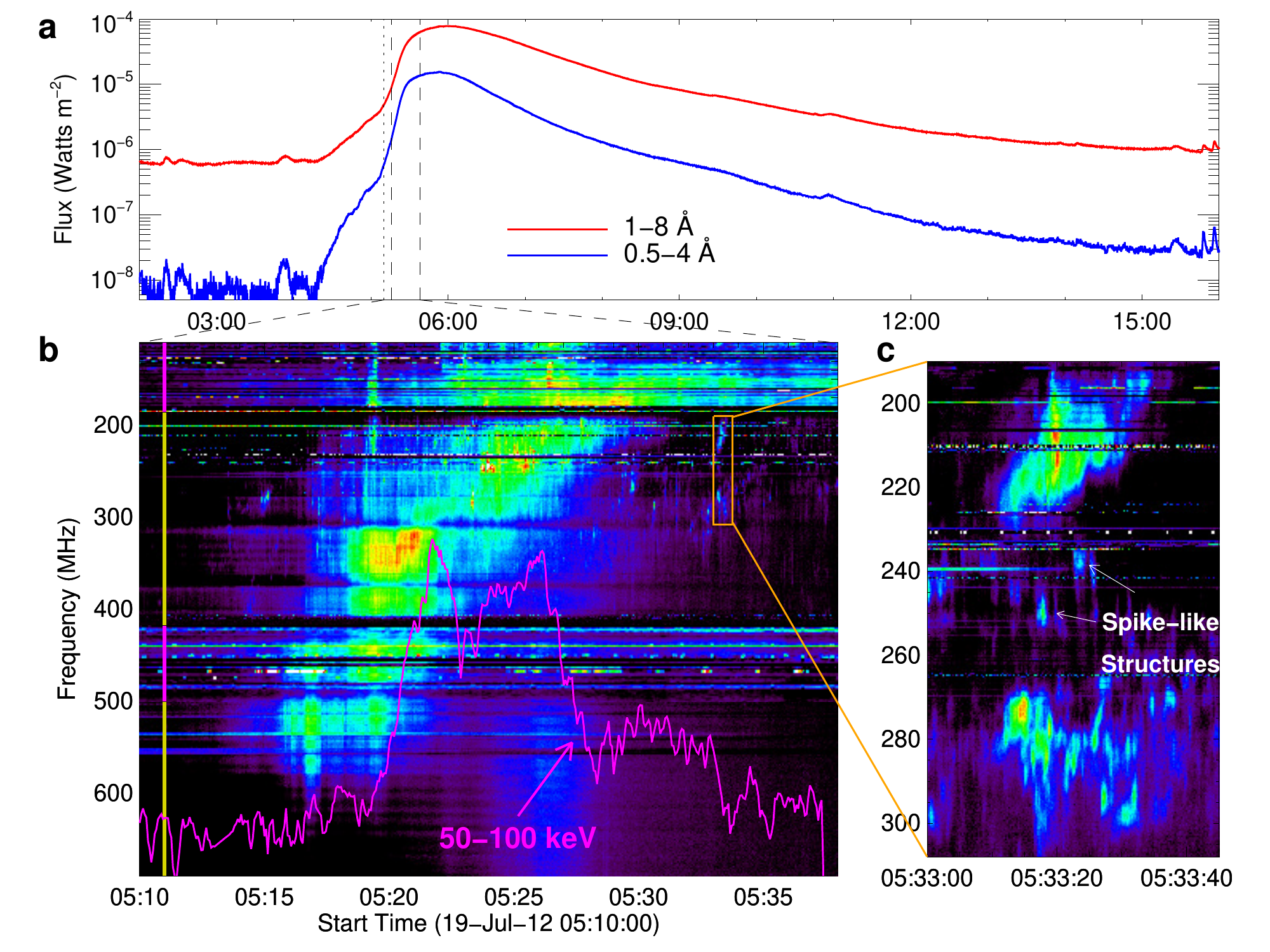}
\caption{The X-ray and radio observations of the flare impulsive phase. (a) GOES soft X-ray fluxes in 1-8 \AA\,(red line) and 0.5-4 \AA\,(blue line), the two vertical dashed lines indicate the flare impulsive phase from 05:16 to 05:43~UT. (b) Combined radio dynamic spectra obtained by the Yunnan (yellow vertical bar) and Culgoora (magenta vertical bar) spectrographs, showing drifting pulsation structures (DPSs). The overplotted magenta and white light curves show the hard X-ray (50-100 keV)  and time derivative of GOES 1-8 {\AA} fluxes, respectively. (c) Zoomed-in view of a pair of oppositely drifting structures. The arrows points to some spike-like structures.
\label{fig:spectrogram}}
\end{figure}

The event of interest with clear observations of a current sheet occurred on July 19, 2012. 
Observations (the 131 {\AA} passband) from the Atmospheric Imaging Assembly (AIA) \citep{Lemen2012} on board Solar Dynamic Observatory (SDO) show  that on July 18  a pre-existing flux rope (Fig.~\ref{fig:blob_img_evo} a) became destabilized, impulsively accelerated (Fig.~\ref{fig:blob_img_evo} b$-$c) and eventually evolved into a white-light CME. 
Behind the CME, a long vertical current sheet is formed, with its lower end connecting to the cusp-shaped flare loops.
According to the peak soft X-ray (1$-$8 \AA) flux (Fig.~\ref{fig:spectrogram} a) recorded by the Geostationary Operational Environmental Satellites (GOES), the flare can be classified as an M7.7 flare. 
 Previous studies have investigated different aspects of the event, for instance, the microwave imaging of the hot flux rope \citep{Wu2016}, a general timeline of the particle acceleration, plasma heating, dynamic processes in the current sheet \citep{Liuw2013, Sun2014, Liur2013, Huang2016}, properties of the above-the-loop-top hard X-ray sources \citep{Krucker2014, Oka2015}, as well as  formation of the related CME \citep{Patsourakos2013}.  Taking advantage of the EUV, X-ray, and radio observations in synergy, here we are moving forward to the most comprehensive and direct observations of plasmoids formed in the current sheet.

The current sheet is found to be highly dynamic.
Starting from about 05:13 UT, a blob-like structure (blob1) was observed to move upward with almost a constant speed of $\sim$ 640 km~s$^{-1}$ (Fig.~\ref{fig:blob_img_evo} d$-$i). Then at about 05:14:57 UT, another  blob (blob2) appeared with an average speed of $\sim$ 1180 km s$^{-1}$ (Fig.~\ref{fig:blob_img_evo} h$-$l). Different from blob1, blob2 moves fast at the beginning, and then slows down.  
Such blobs are interpreted as plasmoids,  which are supposed to be generated due to the tearing instabilities of the continually stretched CS.
After about 05:16 UT, the impulsive phase of the flare starts, showing a sudden increase of hard X-ray (25-50 keV) flux as measured by the Reuven Ramaty High Energy Solar Spectroscopic Imager (RHESSI) \citep{Lin2002}. 
Here the RHESSI thermal (6-12 keV) and nonthermal (25-50 keV) sources (Fig.~\ref{fig:blob_img_evo} m-r)  were reconstructed using the “clean” algorithm and detectors 3, 5, 6, 7, 8, 9. Depending on the count rate, the integration time for the reconstruction ranges from 32 seconds during the flare impulsive phase to 2 minutes during the flare early and decay phases.
Since the production of X-ray bremsstrahlung requires a certain level of target density,  hard X-ray sources are commonly observed at low altitudes at the flare footpoints or above the loop top (cyan contours).
The footpoint sources are generally thought to be produced by  thick-target bremsstrahlung (or braking radiation) of energetic electrons stopped in the cold and dense solar atmosphere \citep{Brown1971,Hudson1972}.
The above-the-loop-top sources from the corona, however, is still under much debate, and several ideas regarding their origin have been proposed \citep{Masuda1994, Fletcher1995, Karlicky2011, Kong2019}. One of them suggests that they can be interpreted as a result of successive merging of plasmoids above the flare loop top, during which a large plasmiod trapping a sufficient amount of energetic particles  is formed \citep{Karlicky2011}.
As revealed in the magnetohydrodynammic and particle kinetic simulations \citep{Kong2020}, the other possibility of the above-the-loop-top sources is the non-thermal electrons accelerated around the up-moving termination shock due to the plasmoid-shock interactions.

\begin{figure}[ht!]
\plotone{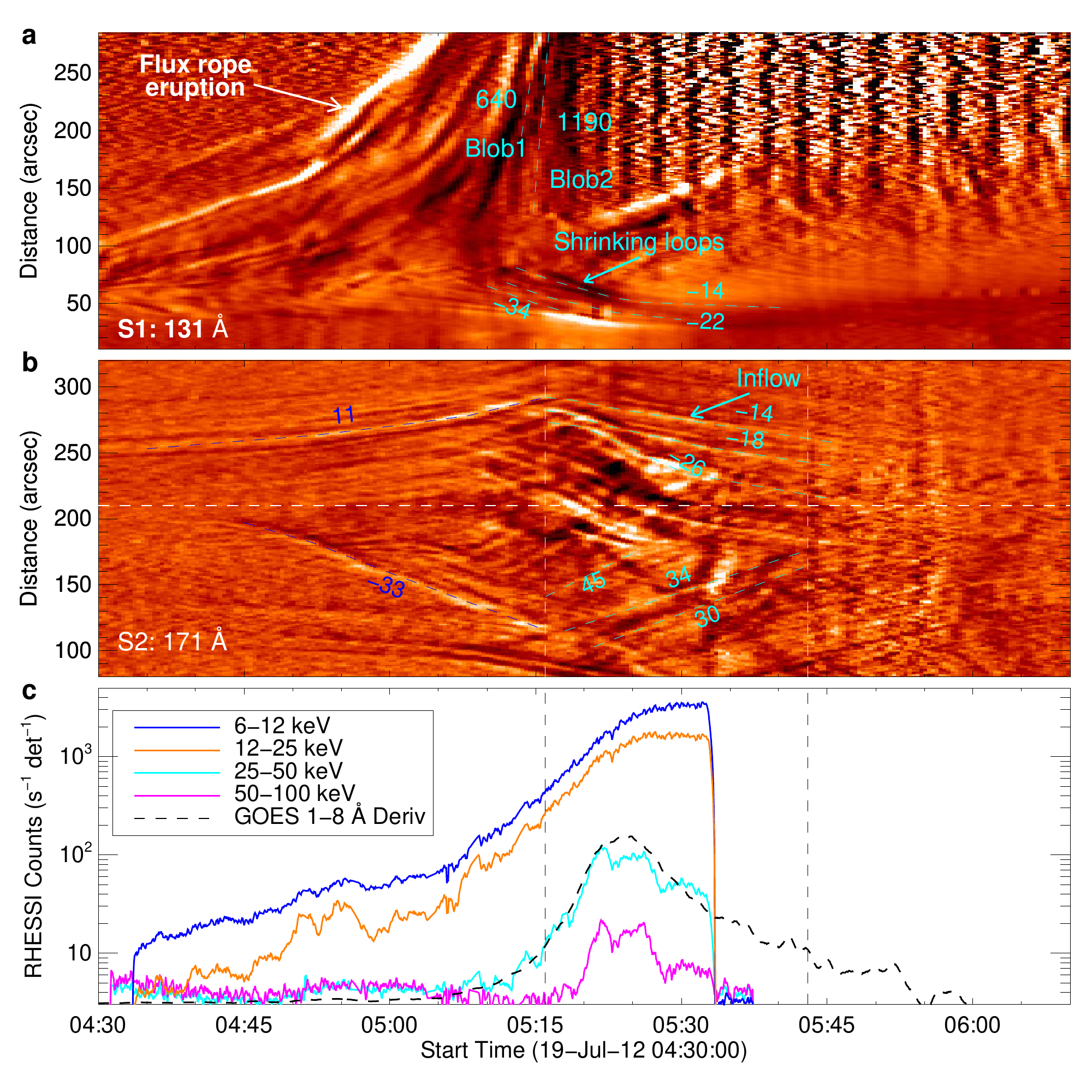}
\caption{Temporal evolution of plasma inflows and outflows during the magnetic reconnection as observed in EUV by SDO/AIA.
 (a) Time-distance plot in AIA 131 \AA ~ along the current sheet (S1), showing the eruption of the flux rope and the bi-directional reconnection outflows. (b) Time-distance plot in AIA 171 \AA ~ along S2 (perpendicular to the current sheet), showing the separation (blue lines) and approach (cyan lines) of oppositely directed magnetic structures (the numbers denote the average speeds in units of km s$^{-1}$). The horizontal white dashed line denotes  the position of the current sheet.
(c) Time profiles of the RHESSI X-ray emission in different energy bands and time derivative of GOES flux in 1-8 {\AA} . The two  vertical dashed lines indicate the flare impulsive phase.  
\label{fig:time-distance}}
\end{figure}

To quantitatively investigate the dynamic process, we select two oblique slices in AIA images (S1 and S2 in Fig. \ref{fig:blob_img_evo} c), with S1 along the CS  and S2 in its perpendicular direction. 
The time-distance plots clearly show that after the eruption of the flux rope,  various plasmoids were formed and ejected along the trailing CS (Fig. \ref{fig:time-distance} a), the ambient coronal plasma that is frozen in the  magnetic field  was first pushed aside (blue lines in Fig. \ref{fig:time-distance} b), then possibly under the restoring force of the magnetic field, the cool plasma (visible in 171 {\AA}, $\sim$ 1 MK) on both sides  keeps converging into the CS with an average velocity of about 16$-$50 km~s$^{-1}$ (cyan lines in Fig. \ref{fig:time-distance} b), and once they come into contact, the magnetic energy pre-stored in the plasma is impulsively released, observed as a sudden flux increase in hard X-ray ($\ge 25$ keV)  
and the time derivative of GOES 1-8 {\AA} flux, according to which the flare impulsive phase is defined (Fig.~\ref{fig:time-distance} c). Note that, the start time of the flare impulsive phase is defined as a sudden rise of the RHESSI 25-50 keV flux while due to the RHESSI night, the end time is defined by the  time derivative of GOES 1-8 {\AA} flux according to the Neupert effect \citep{Neupert1968,Veronig2005}.
Meanwhile, the plasma is strikingly heated and outflow tracers such as fast upward-moving blobs  and downward-shrinking loops become visible in AIA high temperature passband such as 131 {\AA} (cyan lines in Fig. \ref{fig:time-distance} a).

\begin{figure}[ht!]
\plotone{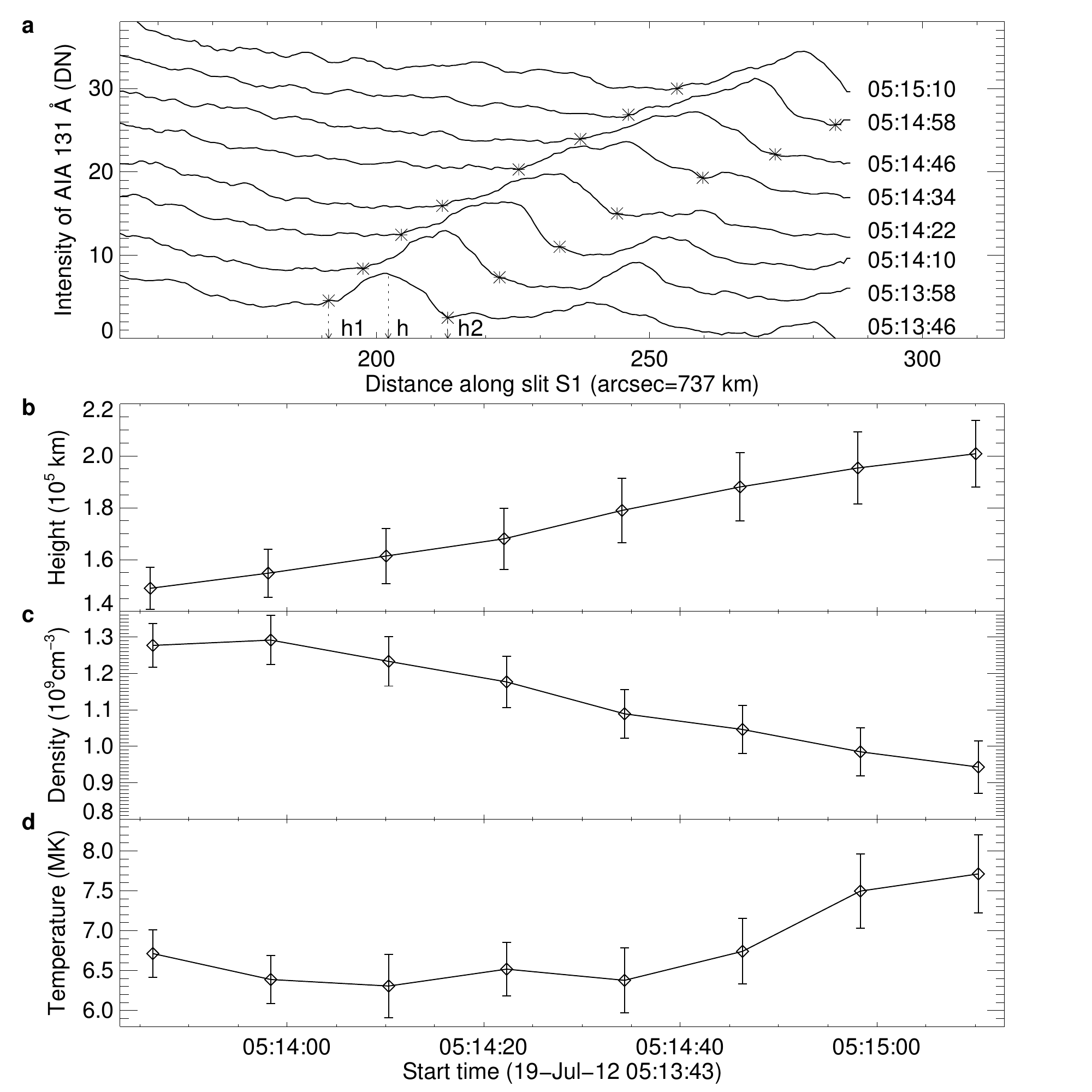}
\caption{Temporal evolution of the  physical parameters of Blob1.
(a) Intensity distributions along S1 at eight selected times. The curves are shifted vertically to avoid overlap. ``h1" and ``h2" indicate the lower and upper boundaries of the blob. 
(b) Evolution of the blob height (h=(h1+h2)/2).  The size (h2-h1) of the blob is regarded as the uncertainties (error bars). 
(c) The electron density inside the blob. (d) The DEM-weighted temperature of the blob. The error bars in (c) and (d) are obtained from 100 Monte Carlo simulations.
\label{fig:blob_par_evo}}
\end{figure}

Fig.~\ref{fig:blob_par_evo} (a) shows the intensity profile of AIA 131 {\AA} along S1 at different times, with h1 and h2 indicating the upper and lower boundaries of blob1. 
The height (h=(h1+h2)/2) and size (s=h2-h1, error bars) of the blob are shown in Fig.~\ref{fig:blob_par_evo} (b).
The average electron density  (Fig~\ref{fig:blob_par_evo} c) as well as  the average  temperature (Fig~\ref{fig:blob_par_evo} d) of the blob was estimated from the analysis of its differential emission measure (DEM) (see Appendix \ref{sec:method_DEM} for details). 
As the blob propagated radially outward, it expanded in size. This expansion is probably due to the ambient magnetic field decreasing with height. Meanwhile, the plasma electron density inside the blob correspondingly decreased. 
The temperature of the blob did not change significantly at the beginning (T $\sim 6.4$ MK), but quickly rose to $7.7$ MK after about 05:15 UT, implying that a number of suprathermal and/or nonthermal electrons may have been injected into the plasmoid and/or the plasmoid is locally energized and heated.

 The plasmoid (blob1) left the field of view (FOV) of AIA at about 05:15:21 UT, and in about two minutes, a clear DPS event (hereafter, the initial DPS),  was observed by Yunnan Astronomical Observatories Radio Spectrometers \citep{Gao2014}.
The DPS consists of many narrow-band pulsation structures with a characteristic repetition time of about 200 seconds \citep{Huang2016}, and as a whole it drifts from high to low frequencies 
at a rate of about -0.68 $\pm$ 0.13 MHz s$^{-1}$,  obtained from a linear fit to its upper and lower boundaries (Fig.~\ref{fig:spectrogram} b).
The magenta and white lines in Fig.~\ref{fig:spectrogram} (b) represent the RHESSI hard X-ray (25-50 keV) time profile and the time derivative of the GOES 1-8 {\AA} light curves, which show a rapid energy release during the DPS.  
Shortly after the initial DPS (from about 05:33:13~UT), a pair of oppositely drifting structures (hereafter, O-DSs)  appeared (Fig.~\ref{fig:spectrogram} c).
The drift rates of the upper and lower branches of the O-DSs were estimated to be -1.96 $\pm$ 0.58 MHz s$^{-1}$ and 1.22 $\pm$ 0.32 MHz s$^{-1}$, respectively. Similar oppositely drifting features  were also reported by \cite{Karlicky2020}. 
According to previous studies \citep{Kliem2000,Shibata2001,Karlicky2002,Karlicky2004}, the initial DPS together with the subsequent O-DSs are supposed to map the evolution of the primary and secondary plasmoids that were formed due to the cascading tearing instabilities in the CS as it was continually stretched. These plasmoids were then exited by energetic electron beams accelerated during the magnetic reconnection process  and producing radio emissions via plasma emission mechanism.  
Moreover, some narrow-band spike-like structures can be distinguished in between the oppositely drifting structures, which is suggestive of the reconnection outflows being in a turbulent state \citep{Barta2001, Karlicky2011,Zhao2018,Jing2019}.

\section{Discussion}

Generally it is assumed that these observed radio frequency $f_{obs}$ is equal to the fundamental or second harmonic of the electron plasma frequency $f_p$ inside a plasmoid, with $f_p$ can be converted into the electron plasma density $n_e$ via $f_p=8980\sqrt{n_e}$. 
The  electron density $n_e$, on the other hand,  is usually related with the height of the plasmoid in the solar corona. 
Thus if we know how the density of the plasmoid varies with its height, we can estimate the kinetics of the radio-related plasmoids.
Here we use a plasmoid density model derived by considering the balance of pressure inside and outside the plasmoid by \cite{Nishizuka2015} (see Appendix \ref{sec:method_density} for details).  
It is found that under the assumption of the second harmonic plasma emission, the physical parameters (height and density) of the plasmoid inferred from the initial DPS are consistent with parameters of the blob1 observed in EUV images considering the two-minute measurement gap. 
Therefore, we suppose that the initial DPS is likely to be generated inside  blob1 or the blob1-liked plasmoid. This could also be supported by the sudden temperature increase of blob1 (implying energetic electron injection) near the early phase of the radio emission.
Based on the  drift rate of the DPS, the velocity of the related plasmoid is estimated to be about 486 km s$^{-1}$, which is smaller than that observed in EUV images,  implying deceleration of blob1 after it left the FOV of AIA.
Again under the assumption of second harmonic plasma emission, the starting heights of the oppositely drifting plasmoids (the O-DSs) were estimated to be  about $3.4 \times 10^5 ~$km and $3.1 \times 10^5 ~$km, respectively. 
At these large heights we are not able to identify the EUV counterparts in AIA images, since the AIA field-of-view is too small. 
The velocity of the downward plasmoid ($\sim$ 508 km s$^{-1}$) is much smaller than that of the upward plasmoid  ($\sim$ 1035 km s$^{-1}$), implying that the downward plasmoids are subject to more resistance than the upward ones.

\section{Summary}

In summary, the multi-waveband and high-resolution observations in EUV,  X-ray, and radio  of the flare allow us to investigate the plasmoids at the vertical current sheet behind an erupting flux rope in great detail. 
The plasmoids were first observed as upward-moving blob-like structures in EUV images during the early phase of the flare. 
Then with the onset of the flare impulsive phase, a large number of electron beams were accelerated, which can excite the plasmoids to oscillate and produce radio emission through plasma emission mechanism or hard X-ray emission via the bremsstrahlung mechanism.
Since energetic electrons require a certain level of target density for the production of nonthermal hard X-ray, hard X-ray sources were only observed near the lower end of the CS (i.e., the above-the-loop-top hard X-ray sources).
Meanwhile a strong DPS, followed by a rare pair of oppositely drifting structures, were clearly observed from radio dynamic spectra.
According to previous studies \citep{Ohyama1998,Kliem2000,Karlicky2004}, these structures were supposed to map the evolution of the primary and secondary plasmoids generated due to the tearing instabilities in the current sheet. 
Based on the observations, we have compared the physical properties of the plasmoids at different stages and found that the initial DPS might evolve from the blob1 seen in AIA images.
The aforementioned analyses show that there are  multiple small-scale short-lived dissipative regions at multiple X lines. Therefore the energy release during the magnetic reconnection occurs probably in a fragmented manner.
These results are consistent with  theoretical prediction of the plasmoid generation by an unstable CS, and support the concept of cascading reconnection.
All these findings help us to obtain a deeper understanding of the magnetic reconnection processes which are the engine of solar eruptions, and shed lights on the energy release in other astrophysics processes.

\begin{acknowledgments}
We thank Hugh Hudson and Jun Lin for carefully reading the paper and for their comments which improved our paper. Also many thanks to Yang Su, Xiaozhou Zhao, Xiangliang Kong,  Xin Cheng, Baolin Tan, Chengming Tan, and Leping Li for helpful discussions. We acknowledge the use of data from SDO, RHESSI, and the solar radio spectrometers of the Yunnan Astronomical Observatories. This work is supported by NSFC (grant Nos. 12103090, U1731241, 11921003, 11973012), the mobility
program (M-0068) of the Sino-German Science Center, by CAS Strategic Pioneer Program on Space Science
(grant Nos., XDA15018300, XDA15052200, XDA15320103, and XDA15320301), and by the National Key R\&D Program
of China (2018YFA0404200). 
L.L. is also supported by CAS Key Laboratory of Solar Activity (KLSA202113).
L.F. also acknowledges the Youth Innovation Promotion Association for financial support. The work of A.W. was supported by the German Space Agency DLR under grant No. 50QL 0001. A.M.V. gratefully acknowledges the support by the Austrian Science Fund (FWF): P27292-N20. 
\end{acknowledgments}

\appendix

\renewcommand{\thefigure}{S\arabic{figure}}
\setcounter{figure}{0}
\renewcommand{\thetable}{S\arabic{table}}
\setcounter{table}{0}

%\section{Appendixes}

In this supplemental material, we provide technical details of main results in the main text.

\section{Differential Emission Measure Reconstruction} \label{sec:method_DEM}

%\label{sec:method_DEM}}

The differential emission measure (DEM) is a physical quantity that measures the amount of  materials emitting at a certain temperature T.
Assuming an optically-thin coronal plasma,  the DEM can be related to the narrow-band EUV (or broadband X-ray) observations $y_i$ as
\begin{equation}
{y_i}=\int K_i(T) ~ {\rm DEM}(T) dT+\delta y_i,
\label{eq:em}
\end{equation}
where $K_i(T)$ is the temperature response function which can be computed using the CHIANTI package\citep{Landi2013}, 
$\delta y_i$ represents a random error that involved in a measurement.
Using the observations in 
AIA's six EUV passbands that are centred on iron
emission lines (94 $\rm \AA$, 131 $\rm \AA$, 171 $\rm \AA$, 193 $\rm \AA$, 211 $\rm \AA$, and 335 $\rm \AA$), the DEM(T) can be inverted. The procedure we use is $xrt\_dem\_iterative2.pro$, which is available in the Solar Software (SSW) package. In the supplementary Fig. \ref{fig:blob_em}, we show 100 Monte Carlo (MC) simulations of the DEM distributions (for each MC, a small random error is added to the observations). The  black line represents the best-fit  DEM distribution.

\begin{figure*}[!h]
\centering
\includegraphics[scale=0.7]{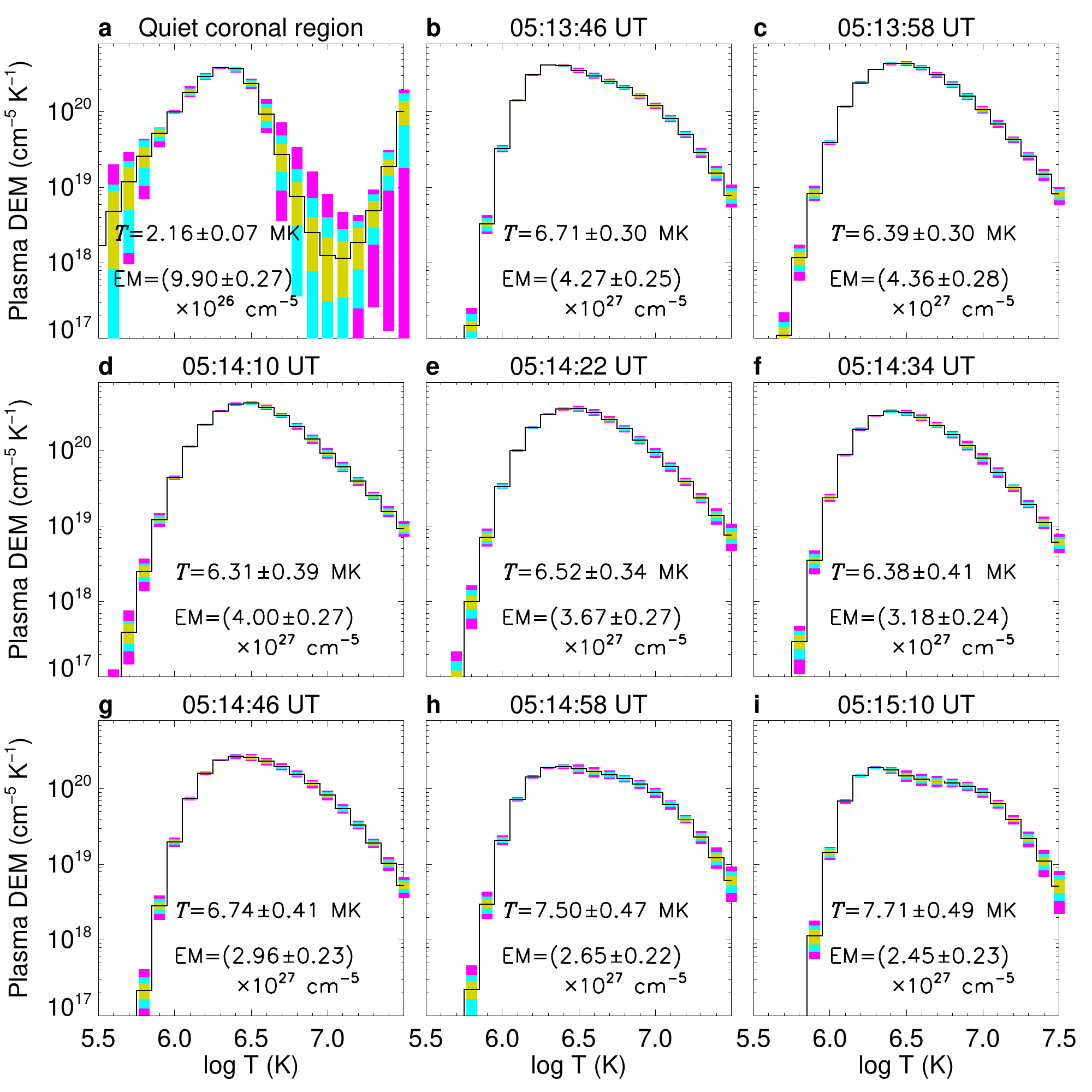}
\caption{Panel (a): DEM of the quiet coronal region (R1 in Fig.~\ref{fig:blob_img_evo}g), Panels (b-i): DEM of the blob region  (R2 in Fig.~\ref{fig:blob_img_evo}g) at eight selected times.
The yellow, cyan and purple colors represent regions in which 50\%, 80\% and 95\% of 100 Monte Carlo (MC) solutions are contained, respectively.
The black line represents the best-fit DEM distribution. The total emission measure and the DEM-weighted temperatures are given in each panel. Three standard deviations (3$\sigma$) of the 100 Monte Carlo simulations are regarded as the uncertainties.
\label{fig:blob_em}}
\end{figure*}

The total emission measure EM  and the DEM-weighted temperature $T_{mean}$ are calculated via the following equations:
\begin{equation}
{\rm EM}=\int_{T_{min}}^{T_{max}} {\rm DEM}(T) dT,
\label{eq:em}
\end{equation}
\begin{equation}
{T_{mean}}=\frac{\int_{T_{min}}^{T_{max}} {\rm DEM}(T) T dT}{\int_{T_{min}}^{T_{max}} {\rm DEM}(T) dT},
\label{eq:temprature}
\end{equation}
The temperature range is set to be 6.0 $\le {\rm log ~}T \le$ 6.5 for the quiet corona  and  $6.0 \le  log T \leq 7.3$ for the plasmoid, within which the DEM is well constrained (see Fig.~S1). 
The average electron plasma density inside the plasmoid is deduced via,
\begin{equation}
n_e=\sqrt{\frac{{\rm EM}}{l}} \quad [{\rm cm^{-3}}]
\label{eq:ne}
\end{equation}
where $l$ represents the effective depth along the line of sight. We assumed that the plasmoid has a similar extent in depth and in length. The length of the plasmoid is shown by the error bars in Fig.~\ref{fig:blob_par_evo}(b), from which we estimate its average value of $\sim 2.4 \times 10^{9}$ cm.
Three standard deviations (3$\sigma$) of 100 Monte Carlo simulations away from the best fit are considered as the uncertainties of the temperature, EM and electron density for the plasmoid (see error bars in Fig.~\ref{fig:blob_par_evo}c-d).

\section{Plasmoid Density Model}
\label{sec:method_density}

The plasmoid density model, i.e., the variance of the electron density inside the plasmoid $n_{in}$ with its height $h$ in the solar corona, is derived by considering the balance of pressure inside and outside the plasmoid by \cite{Nishizuka2015}. The gas and magnetic pressure outside the plasmoid are estimated by assuming a  plasma density model and a dipolar global magnetic field, respectively.
Here we only give the derived expression(see \cite{Nishizuka2015} for details):

%\begin{equation*} 
\[
  n_{in}(h)=\begin{dcases}
              n_{1}(\dfrac{h_{1}}{h})^{p}\dfrac{T_{out}}{T_{in}}+\dfrac{B_{0}^2}{8\pi2k_{B}T_{in}}(1+\dfrac{h}{h_{D}})^{-6}, \quad \quad  (h \leq h_1) \\
               n_{Q}exp(-\dfrac{h}{\lambda_{T}})\dfrac{T_{out}}{T_{in}}+ \dfrac{B_{0}^2}{8\pi2k_{B}T_{in}}(1+\dfrac{h}{h_{D}})^{-6}, 
               \quad \quad (h \geq h_1)
            \end{dcases}
\label{eq:Ne2h}
\]
%\end{equation*}

where $ h_1=1.6 \times 10^{10}\,$~cm is the transition height between the lower and the higher corona, $p = 2.38$ is the power-law index, $ n_1=4.6 \times 10^7\,\rm{cm}^{-3}$,
$ n_Q=4.6 \times 10^8\,\rm{cm}^{-3} $ are the base densities, $k_B=1.38 \times 10^{-23}\,\rm{J/K}$ is the Boltzmann constant, and $\lambda_T=RT/g=6.9 \times 10^9\,\rm{cm}$ is the density scale. 
$B_0 = 100$ G and $h_D = 7.5 \times 10^8$\,~cm are two parameters used to estimate the coronal magnetic field.
$T_{in}$ and $T_{out}$ represent temperatures inside and outside the plasmoid, which are obtained from our DEM results.

The maximal and minimal electron density $n_{max}$ and $n_{min}$ derived from the frequency range of the DPS are then converted into the minimal and maximal heights $h_{min}$ and $h_{max}$ of the plasmoid. Subsequently, the height and size of the plasmoid are obtained from  $h=(h_{max}+h_{min})/2$, and $W_{pla} = h_{max}-h_{min}$.

For each DPS in Fig. \ref{fig:spectrogram}, two time instances are selected for calculation.
The supplementary Table \ref{table:Radio-inferred} shows the physical parameters of the related plasmoids inferred by assuming both fundamental (the bracketed values) and second harmonic plasma emission. As can be seen, comparing to the results with the fundamental plasma emission assumption, the results calculated  by assuming the second harmonic plasma emission  reveal a higher plasma density and smaller coronal height of the related plasmoid.

%%
%% TABLES
%%
%% If there are any tables, put them here.

\begin{table*}
\centering
% title of Table
  \caption{Radio-inferred parameters of the plasmoid.}
  \label{table:Radio-inferred}      % is used to refer this table in the text
\begin{tabular}{|l|c|c|c|c|c|c|}
\hline 
Radio Features& \multicolumn{2}{c|}{the Initial DPS} & \multicolumn{2}{c|}{DS-upper} & \multicolumn{2}{c|}{DS-lower} \\ \hline
Time (UT) & 05:17:00 & 05:24:00 & 05:33:14 & 05:33:20 & 05:33:14 & 05:33:20 \\ \hline
$f_{max}~({\rm MHz})$ & $610$  & $360$  & $230$  & $225$  & $285$   & $293$    \\ \hline
$f_{min}~({\rm MHz})$ & $420$  & $110$  &  $212$ &  $195$ & $267$   & $275$    \\ \hline
$n_{max}~({\rm 10^8~cm^{-3}})$ & $11.5(46.1)$  & $4.02(16.1)$  & $1.64(6.56)$  & $1.57(6.28)$  & $2.52(10.1)$   &   $2.66(10.6)$  \\ \hline
$n_{min}~({\rm 10^8~cm^{-3}})$ & $5.47(21.9)$  & $0.38(1.50)$  & $1.39(5.57)$  & $1.18(4.72)$  &  $2.21(8.84)$   &  $2.34(9.38)$   \\ \hline
$h_{min}~({\rm 10^5~km})$      & $2.41(1.88)$    & $2.95(2.27)$    & $3.34(2.63)$  & $3.36(2.65)$  &  $3.10(2.44)$  &  $3.07(2.42)$   \\ \hline
$h_{max}~({\rm 10^5~km})$      & $2.78(2.15)$    & $6.32(3.67)$    & $3.43(2.71)$  & $3.53(2.78)$  &  $3.17(2.50)$  &  $3.14(2.47)$   \\ \hline   
$h_{pla}~({\rm 10^5~km})$      & $2.59(2.02)$    & $4.64(2.97)$    & $3.38(2.67)$  & $3.44(2.72)$  &  $3.13(2.47)$  & $3.10(2.45)$    \\ \hline               
$W_{pla}~({\rm 10^4~km})$      & $3.62(2.66)$    & $33.7(14.0)$    & $0.94(0.75)$  & $1.68(1.33)$  & $0.70(0.55)$   & $0.67(0.53)$    \\ \hline                
$v_{pla}~({\rm kms^{-1}})$ & \multicolumn{2}{c|}{$486 (227)$} & \multicolumn{2}{c|}{$1035 (822)$} & \multicolumn{2}{c|}{$-508 (-403)$} \\ \hline                 
\end{tabular}
\begin{tablenotes}
 \footnotesize
 \item \textbf{Notes.}``DS-upper'' means the upper branch (drifting towards lower frequencies)  and ``DS-lower'' means the lower branch (drifting towards higher frequencies) of the oppositely drifting structures. $f_{max}$ and $f_{min}$ are the high and low frequency cutoffs. By default, the values are computed by assuming the second harmonic plasma emission. The bracketed values indicate the results calculated with the fundamental plasma emission assumption. 
\end{tablenotes}
 
\end{table*}

\bibliography{mybibfile}{}

\begin{thebibliography}{}
\expandafter\ifx\csname natexlab\endcsname\relax\def\natexlab#1{#1}\fi
\providecommand{\url}[1]{\href{#1}{#1}}
\providecommand{\dodoi}[1]{doi:~\href{http://doi.org/#1}{\nolinkurl{#1}}}
\providecommand{\doeprint}[1]{\href{http://ascl.net/#1}{\nolinkurl{http://ascl.net/#1}}}
\providecommand{\doarXiv}[1]{\href{https://arxiv.org/abs/#1}{\nolinkurl{https://arxiv.org/abs/#1}}}

\bibitem[{{B{\'a}rta} {et~al.}(2011){B{\'a}rta}, {B{\"u}chner}, {Karlick{\'y}},
  \& {Sk{\'a}la}}]{Barta2011}
{B{\'a}rta}, M., {B{\"u}chner}, J., {Karlick{\'y}}, M., \& {Sk{\'a}la}, J.
  2011, \apj, 737, 24, \dodoi{10.1088/0004-637X/737/1/24}

\bibitem[{{B{\'a}rta} \& {Karlick{\'y}}(2001)}]{Barta2001}
{B{\'a}rta}, M., \& {Karlick{\'y}}, M. 2001, \aap, 379, 1045,
  \dodoi{10.1051/0004-6361:20011375}

\bibitem[{{Bhattacharjee} {et~al.}(2009){Bhattacharjee}, {Huang}, {Yang}, \&
  {Rogers}}]{Bhattacharjee2009}
{Bhattacharjee}, A., {Huang}, Y.-M., {Yang}, H., \& {Rogers}, B. 2009, Physics
  of Plasmas, 16, 112102, \dodoi{10.1063/1.3264103}

\bibitem[{{Brown}(1971)}]{Brown1971}
{Brown}, J.~C. 1971, \solphys, 18, 489, \dodoi{10.1007/BF00149070}

\bibitem[{{Cairns} {et~al.}(2018){Cairns}, {Lobzin}, {Donea}, {Tingay},
  {McCauley}, {Oberoi}, {Duffin}, {Reiner}, {Hurley-Walker}, {Kudryavtseva},
  {Melrose}, {Harding}, {Bernardi}, {Bowman}, {Cappallo}, {Corey}, {Deshpand
  e}, {Emrich}, {Goeke}, {Hazelton}, {Johnston-Hollitt}, {Kaplan}, {Kasper},
  {Kratzenberg}, {Lonsdale}, {Lynch}, {McWhirter}, {Mitchell}, {Morales},
  {Morgan}, {Ord}, {Prabu}, {Roshi}, {Shankar}, {Srivani}, {Subrahmanyan},
  {Wayth}, {Waterson}, {Webster}, {Whitney}, {Williams}, \&
  {Williams}}]{Cairns2018}
{Cairns}, I.~H., {Lobzin}, V.~V., {Donea}, A., {et~al.} 2018, Scientific
  Reports, 8, 1676, \dodoi{10.1038/s41598-018-19195-3}

\bibitem[{{Carmichael}(1964)}]{Carmichael1964}
{Carmichael}, H. 1964, NASA Special Publication, 50, 451

\bibitem[{{Dong} {et~al.}(2012){Dong}, {Wang}, {Lu}, {Huang}, {Yuan}, {Liu},
  {Lin}, {Li}, {Wei}, {Zhong}, {Shi}, {Jiang}, {Ding}, {Jiang}, {Du}, {He},
  {Yu}, {Liu}, {Wang}, {Tang}, {Zhu}, {Zhao}, {Sheng}, \& {Zhang}}]{Dong2012}
{Dong}, Q.-L., {Wang}, S.-J., {Lu}, Q.-M., {et~al.} 2012, \prl, 108, 215001,
  \dodoi{10.1103/PhysRevLett.108.215001}

\bibitem[{{Fletcher}(1995)}]{Fletcher1995}
{Fletcher}, L. 1995, \aap, 303, L9

\bibitem[{{Fletcher}(2005)}]{Fletcher2005}
---. 2005, \ssr, 121, 141, \dodoi{10.1007/s11214-006-7181-7}

\bibitem[{{Gao} {et~al.}(2014){Gao}, {Wang}, {Dong}, {Wu}, \& {Lin}}]{Gao2014}
{Gao}, G., {Wang}, M., {Dong}, L., {Wu}, N., \& {Lin}, J. 2014, \na, 30, 68,
  \dodoi{10.1016/j.newast.2014.01.008}

\bibitem[{{Gou} {et~al.}(2019){Gou}, {Liu}, {Kliem}, {Wang}, \&
  {Veronig}}]{Gou2019}
{Gou}, T., {Liu}, R., {Kliem}, B., {Wang}, Y., \& {Veronig}, A.~M. 2019,
  Science Advances, 5, 7004, \dodoi{10.1126/sciadv.aau7004}

\bibitem[{{Hirayama}(1974)}]{Hirayama1974}
{Hirayama}, T. 1974, \solphys, 34, 323, \dodoi{10.1007/BF00153671}

\bibitem[{{Huang} {et~al.}(2016){Huang}, {Kontar}, {Nakariakov}, \&
  {Gao}}]{Huang2016}
{Huang}, J., {Kontar}, E.~P., {Nakariakov}, V.~M., \& {Gao}, G. 2016, \apj,
  831, 119, \dodoi{10.3847/0004-637X/831/2/119}

\bibitem[{{Huang} \& {Bhattacharjee}(2010)}]{Huang2010}
{Huang}, Y.-M., \& {Bhattacharjee}, A. 2010, Physics of Plasmas, 17, 062104,
  \dodoi{10.1063/1.3420208}

\bibitem[{{Hudson}(1972)}]{Hudson1972}
{Hudson}, H.~S. 1972, \solphys, 24, 414, \dodoi{10.1007/BF00153384}

\bibitem[{{Hudson} {et~al.}(2001){Hudson}, {Kosugi}, {Nitta}, \&
  {Shimojo}}]{Hudson2001}
{Hudson}, H.~S., {Kosugi}, T., {Nitta}, N.~V., \& {Shimojo}, M. 2001, \apjl,
  561, L211, \dodoi{10.1086/324760}

\bibitem[{{Karlick{\'y}}(2004)}]{Karlicky2004}
{Karlick{\'y}}, M. 2004, \aap, 417, 325, \dodoi{10.1051/0004-6361:20034249}

\bibitem[{{Karlick{\'y}} \& {B{\'a}rta}(2011)}]{Karlicky2011}
{Karlick{\'y}}, M., \& {B{\'a}rta}, M. 2011, \apj, 733, 107,
  \dodoi{10.1088/0004-637X/733/2/107}

\bibitem[{{Karlick{\'y}} {et~al.}(2002){Karlick{\'y}}, {F{\'a}rn{\'{\i}}k}, \&
  {M{\'e}sz{\'a}rosov{\'a}}}]{Karlicky2002}
{Karlick{\'y}}, M., {F{\'a}rn{\'{\i}}k}, F., \& {M{\'e}sz{\'a}rosov{\'a}}, H.
  2002, \aap, 395, 677, \dodoi{10.1051/0004-6361:20021310}

\bibitem[{{Karlick{\'y}} \& {Ryb{\'a}k}(2020)}]{Karlicky2020}
{Karlick{\'y}}, M., \& {Ryb{\'a}k}, J. 2020, \apjs, 250, 31,
  \dodoi{10.3847/1538-4365/abb19f}

\bibitem[{{Kliem} {et~al.}(2000){Kliem}, {Karlick{\'y}}, \& {Benz}}]{Kliem2000}
{Kliem}, B., {Karlick{\'y}}, M., \& {Benz}, A.~O. 2000, \aap, 360, 715

\bibitem[{{Ko} {et~al.}(2003){Ko}, {Raymond}, {Lin}, {Lawrence}, {Li}, \&
  {Fludra}}]{Ko2003}
{Ko}, Y.-K., {Raymond}, J.~C., {Lin}, J., {et~al.} 2003, \apj, 594, 1068,
  \dodoi{10.1086/376982}

\bibitem[{{Kong} {et~al.}(2020){Kong}, {Guo}, {Shen}, {Chen}, {Chen}, \&
  {Giacalone}}]{Kong2020}
{Kong}, X., {Guo}, F., {Shen}, C., {et~al.} 2020, \apjl, 905, L16,
  \dodoi{10.3847/2041-8213/abcbf5}

\bibitem[{{Kong} {et~al.}(2019){Kong}, {Guo}, {Shen}, {Chen}, {Chen}, {Musset},
  {Glesener}, {Pongkitiwanichakul}, \& {Giacalone}}]{Kong2019}
---. 2019, \apjl, 887, L37, \dodoi{10.3847/2041-8213/ab5f67}

\bibitem[{{Kopp} \& {Pneuman}(1976)}]{Kopp1976}
{Kopp}, R.~A., \& {Pneuman}, G.~W. 1976, \solphys, 50, 85,
  \dodoi{10.1007/BF00206193}

\bibitem[{{Krucker} \& {Battaglia}(2014)}]{Krucker2014}
{Krucker}, S., \& {Battaglia}, M. 2014, \apj, 780, 107,
  \dodoi{10.1088/0004-637X/780/1/107}

\bibitem[{{Krucker} {et~al.}(2008){Krucker}, {Battaglia}, {Cargill},
  {Fletcher}, {Hudson}, {MacKinnon}, {Masuda}, {Sui}, {Tomczak}, {Veronig},
  {Vlahos}, \& {White}}]{Krucker2008}
{Krucker}, S., {Battaglia}, M., {Cargill}, P.~J., {et~al.} 2008, \aapr, 16,
  155, \dodoi{10.1007/s00159-008-0014-9}

\bibitem[{{Landi} {et~al.}(2013){Landi}, {Young}, {Dere}, {Del Zanna}, \&
  {Mason}}]{Landi2013}
{Landi}, E., {Young}, P.~R., {Dere}, K.~P., {Del Zanna}, G., \& {Mason}, H.~E.
  2013, \apj, 763, 86, \dodoi{10.1088/0004-637X/763/2/86}

\bibitem[{{Lemen} {et~al.}(2012){Lemen}, {Title}, {Akin}, {Boerner}, {Chou},
  {Drake}, {Duncan}, {Edwards}, {Friedlaender}, {Heyman}, {Hurlburt}, {Katz},
  {Kushner}, {Levay}, {Lindgren}, {Mathur}, {McFeaters}, {Mitchell}, {Rehse},
  {Schrijver}, {Springer}, {Stern}, {Tarbell}, {Wuelser}, {Wolfson}, {Yanari},
  {Bookbinder}, {Cheimets}, {Caldwell}, {Deluca}, {Gates}, {Golub}, {Park},
  {Podgorski}, {Bush}, {Scherrer}, {Gummin}, {Smith}, {Auker}, {Jerram},
  {Pool}, {Soufli}, {Windt}, {Beardsley}, {Clapp}, {Lang}, \&
  {Waltham}}]{Lemen2012}
{Lemen}, J.~R., {Title}, A.~M., {Akin}, D.~J., {et~al.} 2012, \solphys, 275,
  17, \dodoi{10.1007/s11207-011-9776-8}

\bibitem[{{Li}(2019)}]{Lid2019}
{Li}, D. 2019, Research in Astronomy and Astrophysics, 19, 067,
  \dodoi{10.1088/1674-4527/19/5/67}

\bibitem[{{Li} {et~al.}(2016){Li}, {Zhang}, {Peter}, {Priest}, {Chen}, {Guo},
  {Chen}, \& {Mackay}}]{Li2016}
{Li}, L., {Zhang}, J., {Peter}, H., {et~al.} 2016, Nature Physics, 12, 847,
  \dodoi{10.1038/nphys3768}

\bibitem[{{Lin} {et~al.}(2008){Lin}, {Cranmer}, \& {Farrugia}}]{Lin2008}
{Lin}, J., {Cranmer}, S.~R., \& {Farrugia}, C.~J. 2008, Journal of Geophysical
  Research (Space Physics), 113, A11107, \dodoi{10.1029/2008JA013409}

\bibitem[{{Lin} {et~al.}(2002){Lin}, {Dennis}, {Hurford}, {Smith}, {Zehnder},
  {Harvey}, {Curtis}, {Pankow}, {Turin}, {Bester}, {Csillaghy}, {Lewis},
  {Madden}, {van Beek}, {Appleby}, {Raudorf}, {McTiernan}, {Ramaty}, {Schmahl},
  {Schwartz}, {Krucker}, {Abiad}, {Quinn}, {Berg}, {Hashii}, {Sterling},
  {Jackson}, {Pratt}, {Campbell}, {Malone}, {Landis}, {Barrington-Leigh},
  {Slassi-Sennou}, {Cork}, {Clark}, {Amato}, {Orwig}, {Boyle}, {Banks},
  {Shirey}, {Tolbert}, {Zarro}, {Snow}, {Thomsen}, {Henneck}, {McHedlishvili},
  {Ming}, {Fivian}, {Jordan}, {Wanner}, {Crubb}, {Preble}, {Matranga}, {Benz},
  {Hudson}, {Canfield}, {Holman}, {Crannell}, {Kosugi}, {Emslie}, {Vilmer},
  {Brown}, {Johns-Krull}, {Aschwanden}, {Metcalf}, \& {Conway}}]{Lin2002}
{Lin}, R.~P., {Dennis}, B.~R., {Hurford}, G.~J., {et~al.} 2002, \solphys, 210,
  3, \dodoi{10.1023/A:1022428818870}

\bibitem[{{Liu}(2013)}]{Liur2013}
{Liu}, R. 2013, \mnras, 434, 1309, \dodoi{10.1093/mnras/stt1090}

\bibitem[{{Liu} {et~al.}(2010){Liu}, {Lee}, {Wang}, {Stenborg}, {Liu}, \&
  {Wang}}]{Liu2010}
{Liu}, R., {Lee}, J., {Wang}, T., {et~al.} 2010, \apjl, 723, L28,
  \dodoi{10.1088/2041-8205/723/1/L28}

\bibitem[{{Liu} {et~al.}(2013){Liu}, {Chen}, \& {Petrosian}}]{Liuw2013}
{Liu}, W., {Chen}, Q., \& {Petrosian}, V. 2013, \apj, 767, 168,
  \dodoi{10.1088/0004-637X/767/2/168}

\bibitem[{{Loureiro} {et~al.}(2007){Loureiro}, {Schekochihin}, \&
  {Cowley}}]{Loureiro2007}
{Loureiro}, N.~F., {Schekochihin}, A.~A., \& {Cowley}, S.~C. 2007, Physics of
  Plasmas, 14, 100703, \dodoi{10.1063/1.2783986}

\bibitem[{{Masuda} {et~al.}(1994){Masuda}, {Kosugi}, {Hara}, {Tsuneta}, \&
  {Ogawara}}]{Masuda1994}
{Masuda}, S., {Kosugi}, T., {Hara}, H., {Tsuneta}, S., \& {Ogawara}, Y. 1994,
  \nat, 371, 495, \dodoi{10.1038/371495a0}

\bibitem[{{Mei} {et~al.}(2012){Mei}, {Shen}, {Wu}, {Lin}, {Murphy}, \&
  {Roussev}}]{Mei2012}
{Mei}, Z., {Shen}, C., {Wu}, N., {et~al.} 2012, \mnras, 425, 2824,
  \dodoi{10.1111/j.1365-2966.2012.21625.x}

\bibitem[{{Mei} {et~al.}(2017){Mei}, {Keppens}, {Roussev}, \& {Lin}}]{Mei2017}
{Mei}, Z.~X., {Keppens}, R., {Roussev}, I.~I., \& {Lin}, J. 2017, \aap, 604,
  L7, \dodoi{10.1051/0004-6361/201731146}

\bibitem[{{Neupert}(1968)}]{Neupert1968}
{Neupert}, W.~M. 1968, \apjl, 153, L59, \dodoi{10.1086/180220}

\bibitem[{{Ni} {et~al.}(2015){Ni}, {Kliem}, {Lin}, \& {Wu}}]{Ni2015}
{Ni}, L., {Kliem}, B., {Lin}, J., \& {Wu}, N. 2015, \apj, 799, 79,
  \dodoi{10.1088/0004-637X/799/1/79}

\bibitem[{{Nishizuka} {et~al.}(2015){Nishizuka}, {Karlick{\'y}}, {Janvier}, \&
  {B{\'a}rta}}]{Nishizuka2015}
{Nishizuka}, N., {Karlick{\'y}}, M., {Janvier}, M., \& {B{\'a}rta}, M. 2015,
  \apj, 799, 126, \dodoi{10.1088/0004-637X/799/2/126}

\bibitem[{{Ohyama} \& {Shibata}(1998)}]{Ohyama1998}
{Ohyama}, M., \& {Shibata}, K. 1998, \apj, 499, 934, \dodoi{10.1086/305652}

\bibitem[{{Oka} {et~al.}(2015){Oka}, {Krucker}, {Hudson}, \&
  {Saint-Hilaire}}]{Oka2015}
{Oka}, M., {Krucker}, S., {Hudson}, H.~S., \& {Saint-Hilaire}, P. 2015, \apj,
  799, 129, \dodoi{10.1088/0004-637X/799/2/129}

\bibitem[{{Patsourakos} {et~al.}(2013){Patsourakos}, {Vourlidas}, \&
  {Stenborg}}]{Patsourakos2013}
{Patsourakos}, S., {Vourlidas}, A., \& {Stenborg}, G. 2013, \apj, 764, 125,
  \dodoi{10.1088/0004-637X/764/2/125}

\bibitem[{{Priest} \& {Forbes}(2000)}]{Priest2000}
{Priest}, E., \& {Forbes}, T. 2000, {Magnetic Reconnection} (Cambridge
  University Press, Cambridge), 612

\bibitem[{{Samtaney} {et~al.}(2009){Samtaney}, {Loureiro}, {Uzdensky},
  {Schekochihin}, \& {Cowley}}]{Samtaney2009}
{Samtaney}, R., {Loureiro}, N.~F., {Uzdensky}, D.~A., {Schekochihin}, A.~A., \&
  {Cowley}, S.~C. 2009, \prl, 103, 105004,
  \dodoi{10.1103/PhysRevLett.103.105004}

\bibitem[{{Shen} {et~al.}(2011){Shen}, {Lin}, \& {Murphy}}]{Shen2011}
{Shen}, C., {Lin}, J., \& {Murphy}, N.~A. 2011, \apj, 737, 14,
  \dodoi{10.1088/0004-637X/737/1/14}

\bibitem[{{Shibata} \& {Magara}(2011)}]{Shibata2011}
{Shibata}, K., \& {Magara}, T. 2011, Living Reviews in Solar Physics, 8, 6,
  \dodoi{10.12942/lrsp-2011-6}

\bibitem[{{Shibata} {et~al.}(1995){Shibata}, {Masuda}, {Shimojo}, {Hara},
  {Yokoyama}, {Tsuneta}, {Kosugi}, \& {Ogawara}}]{Shibata1995}
{Shibata}, K., {Masuda}, S., {Shimojo}, M., {et~al.} 1995, \apjl, 451, L83,
  \dodoi{10.1086/309688}

\bibitem[{{Shibata} \& {Tanuma}(2001)}]{Shibata2001}
{Shibata}, K., \& {Tanuma}, S. 2001, Earth, Planets, and Space, 53, 473,
  \dodoi{10.1186/BF03353258}

\bibitem[{{Shimojo} {et~al.}(2017){Shimojo}, {Hudson}, {White}, {Bastian}, \&
  {Iwai}}]{Shimojo2017}
{Shimojo}, M., {Hudson}, H.~S., {White}, S.~M., {Bastian}, T.~S., \& {Iwai}, K.
  2017, \apj, 841, L5, \dodoi{10.3847/2041-8213/aa70e3}

\bibitem[{{Sturrock}(1966)}]{Sturrock1966}
{Sturrock}, P.~A. 1966, \nat, 211, 695, \dodoi{10.1038/211695a0}

\bibitem[{{Su} {et~al.}(2013){Su}, {Veronig}, {Holman}, {Dennis}, {Wang},
  {Temmer}, \& {Gan}}]{Su2013}
{Su}, Y., {Veronig}, A.~M., {Holman}, G.~D., {et~al.} 2013, Nature Physics, 9,
  489, \dodoi{10.1038/nphys2675}

\bibitem[{{Sun} {et~al.}(2014){Sun}, {Cheng}, \& {Ding}}]{Sun2014}
{Sun}, J.~Q., {Cheng}, X., \& {Ding}, M.~D. 2014, \apj, 786, 73,
  \dodoi{10.1088/0004-637X/786/1/73}

\bibitem[{{Uzdensky} {et~al.}(2010){Uzdensky}, {Loureiro}, \&
  {Schekochihin}}]{Uzdensky2010}
{Uzdensky}, D.~A., {Loureiro}, N.~F., \& {Schekochihin}, A.~A. 2010, Physical
  Review Letters, 105, 235002, \dodoi{10.1103/PhysRevLett.105.235002}

\bibitem[{{Veronig} {et~al.}(2005){Veronig}, {Brown}, {Dennis}, {Schwartz},
  {Sui}, \& {Tolbert}}]{Veronig2005}
{Veronig}, A.~M., {Brown}, J.~C., {Dennis}, B.~R., {et~al.} 2005, \apj, 621,
  482, \dodoi{10.1086/427274}

\bibitem[{{Wang} {et~al.}(2010){Wang}, {Lu}, {Du}, \& {Wang}}]{Wang2010}
{Wang}, R., {Lu}, Q., {Du}, A., \& {Wang}, S. 2010, \prl, 104, 175003,
  \dodoi{10.1103/PhysRevLett.104.175003}

\bibitem[{{Wu} {et~al.}(2016){Wu}, {Chen}, {Huang}, {Nakajima}, {Song},
  {Melnikov}, {Liu}, {Li}, {Chandrashekhar}, \& {Jiao}}]{Wu2016}
{Wu}, Z., {Chen}, Y., {Huang}, G., {et~al.} 2016, \apjl, 820, L29,
  \dodoi{10.3847/2041-8205/820/2/L29}

\bibitem[{{Ye} {et~al.}(2019){Ye}, {Shen}, {Raymond}, {Lin}, \&
  {Ziegler}}]{Jing2019}
{Ye}, J., {Shen}, C., {Raymond}, J.~C., {Lin}, J., \& {Ziegler}, U. 2019,
  \mnras, 482, 588, \dodoi{10.1093/mnras/sty2716}

\bibitem[{{Zhao} {et~al.}(2018){Zhao}, {Ni}, {Lin}, \& {Ziegler}}]{Zhao2018}
{Zhao}, T.-L., {Ni}, L., {Lin}, J., \& {Ziegler}, U. 2018, Research in
  Astronomy and Astrophysics, 18, 045, \dodoi{10.1088/1674-4527/18/4/45}

\bibitem[{{Zhao} {et~al.}(2021){Zhao}, {Bacchini}, \& {Keppens}}]{Zhao2021}
{Zhao}, X., {Bacchini}, F., \& {Keppens}, R. 2021, Physics of Plasmas, 28,
  092113, \dodoi{10.1063/5.0058326}

\bibitem[{{Zweibel} \& {Yamada}(2009)}]{Zweibel2009}
{Zweibel}, E.~G., \& {Yamada}, M. 2009, \araa, 47, 291,
  \dodoi{10.1146/annurev-astro-082708-101726}

\end{thebibliography}
\bibliographystyle{aasjournal}

%% This command is needed to show the entire author+affiliation list when
%% the collaboration and author truncation commands are used.  It has to
%% go at the end of the manuscript.
%\allauthors

%% Include this line if you are using the \added, \replaced, \deleted
%% commands to see a summary list of all changes at the end of the article.
%\listofchanges

\end{document}